\documentclass{article}

\begin{document}

\title{Two effective temperatures in traffic flow models: analogies with granular
flow}
\author{M. E. L\'{a}rraga, J. A. del R\'{\i}o \\
Centro de Investigaci\'{o}n en Energ\'{\i}a,\\
Universidad Nacional Aut\'{o}noma de M\'{e}xico,\\
A.P.34, 62580 Temixco, Mor. M\'{e}xico\\
email: mel@cie.unam.mx, antonio@servidor.unam.mx \and Anita Mehta \\
S N Bose National Centre for Basic Sciences, \\
Block JD, Sector III, Salt Lake, Calcutta 700 098,\\
India, \\
email: anita@boson.bose.res.in}
\maketitle

\begin{abstract}
We present a model of traffic flow, with rules that describe the behaviour
of \textit{automated vehicles} in an \textit{open system}. We show first of
all that the fundamental diagram of this system collapses to a point, where
states of free and jammed traffic can coexist in phase space. This leads us
to consider separately the roles of average velocities and densities as
better descriptors of the actual state of the traffic. Next, we observe that
the transition between free and jammed traffic as a function of the \textit{%
braking parameter} R is different for high and low initial densities, in the
steady state; it turns out to be 'smeared out' for low densities, a
behaviour which is already portended by the transient behaviour of the
system. Our results indicate strongly that, at least for such models, two
effective temperatures (one related to R, and the other to the density) are
needed to describe the global behaviour of this system in statistical
mechanical terms. Analogies with granular flow are discussed in this context.
\end{abstract}

\section{Introduction}

Traffic flow\cite{tft} is a subject of interdisciplinary\cite{pr} interest
at the present time, both to do with the very real problems of congestion on
busy highways\cite{ril}, as well as an example of a complex system whose
behaviour has yet to be fully understood. The study of this system started
as early as 1934 by Greenshields with a study on traffic capacity\cite{green}%
. Since its inception, the study of traffic flow was based on stochastic
processes\cite{adams}. In 1955 Lighthill and Witham described the existence
of density waves as well as shock waves\cite{lighthill} in a continuous
model of traffic flow. This was also obtained as a limiting case from a
Boltzmann-like model by Prigogine and co-workers\cite{prigogine}. In these
models, traffic flow was treated in an average sense. The discussion on the
''temperature'' of this kind of systems has also been discussed from the
thermodynamic point of view\cite{jsp}. On the other other hand,
car-following theories deal with equations of motion for individual cars, by
using the analogues of Newton's equations of motion\cite{pr}. As early as
1958, Chandler et al. \cite{chandler} have argued that one should take into
account the reaction times of drivers, in response to the cars in front of
them. From this ''microscopic'' perspective one obtains nonlinear
differential equations which describe flow-density relations characteristic
of current models \cite{nagelpre}. In 1956 Gerlough proposed a cellular
automaton model for traffic flow; similar models have been the subject of
recent work\cite{nagel}. Finally, analogies between vehicular traffic with
granular flow, while longstanding, \cite{leutzbach} have also been the
subject of much current interest\cite{wolf},\cite{book},\cite{book98}. In
this paper we concern ourselves largely with these two latter aspects.

In this paper, we focus on a new and rather specific analogy between traffic
and granular flow \cite{book} which concerns the role of temperature; the
results of our model suggest that \textit{two} effective temperatures are
needed to describe the flow of traffic in thermodynamic terms, which is
analogous to the situation that obtains in models of shaken sand \cite{paj}.

We have modified\cite{mariaelena} the extensively studied model of Nagel and
Schreckenberg (NaSch)\cite{pr},\cite{nagel}, \cite{ss} to consider \textit{%
automated vehicles}. This was motivated by a realistic system, which is a
car equipped with an infrared sensor to determine distance and velocity from
its neighbouring vehicles: this was recently used by the Intelligent
Transport Society to conduct studies on automated vehicles \cite{ril}. The
objective of our model is\ also to study the effect of open boundary
conditions (as occurs in reality) on vehicular traffic. In our specific
case, this involves four cell-updating steps involving braking, stochastic
driver reaction, and car movement/acceleration. The chief modification we
have made to the NaSch model involves stochastic changes to the car
occurring \textit{before} the braking step, to model the behaviour of an
automated car or an 'anticipatory' driver.

In open systems, nonequilibrium conditions greatly modify the underlying
physics; even in the steady state, we find that the construction of the
phase diagram is totally different, involving as it does a collapse of the
phase space, relative to the so-called 'fundamental' diagram obtained for
the closed version\cite{eis}, \cite{nagel}. This necessitates a description
of the traffic in terms of densities and velocities, since the flux is no
longer a good variable for this open system.

Our results indicate that for high values of the initial density, there is a
sharp transition between free and jammed flow at a certain value $R_{c}$ of
a parameter $R$ (see below) which we call the 'braking parameter' of the
system. This transition gets smeared out, in the case of lower densities, so
that it is possible to get free flow \textit{even} at values of $R\gg R_{c}$%
. We find that both $R$ and the density are needed to describe the behaviour
of the system in thermodynamic terms, just as in the case of shaken powders.
In terms of the analogy with granular flow\cite{snagelreview}, the parameter 
$R$ is found to be related (inversely) to the shaking intensity, and is like
a 'fast dynamics' temperature, while the density is like a 'slow dynamics'
temperature, much like the Edwards' compactivity \cite{sam} for granular
systems. We will give an explanation of this in terms of individual and
collective effects below, as has already been done in the case of granular
flow \cite{mybook}. This is borne out by the study of velocity correlation
functions, at specific values of $R$, which manifest a 'dynamical
clustering' as observed in models of vibrated sandpiles \cite{bm}.

This paper is organized as follows. In the first section we present our
model. In the next section we present results concerning both transient and
steady-state regimes in the open systems under study; we also include a
comparison with a system with periodic boundary conditions. Lastly, we
summarise our observations, and compare our predictions with observations on
real traffic.

\section{The model}

Most studies involving cellular automata modelling of traffic have sought to
focus on its evolution in closed systems, subject to periodic boundary
conditions. However, for both practical and theoretical reasons, open
boundary conditions are preferable. As we all know, traffic flow in the real
world always occurs in open systems, i.e. those where cars are interchanged
between some local environment and its surroundings; thus for example, the
number of cars is \textit{not }conserved in general in any section of a
highway. In particular a one-dimensional example of an open system could be,
for example, a situation where a multilane road is reduced to one lane, e.g.
due to road construction. Recently, a study of open boundary conditions has
been performed characterizing the phase transition from free to jammed flow
in terms of the input and output rates of cars \cite{santen}. What we seek
to do in this study is to extend this line of research, and model in
particular the specific features which obtain when automated vehicles are
considered.

We base our model on the NaSch \cite{nagel} cellular automaton model; our
most important modification involves changing the order of the operators and
an amendment made to the braking step. Our model is defined as follows:

Consider a one-dimensional array of \ $L$ cells. Each cell can either be
empty or occupied by one car with velocity $v$ $\in \{v_{\min },..,\nu
_{\max }\}$ (with $v_{\min }=1$ and $v_{\max }=5)$. The state of car $j$ $%
(j=1,...,N)$ is characterized by an internal parameter $v_{j}$ $%
(v_{j}=1,...,v_{\max })$, the instantaneous velocity of the vehicle. Let $%
x_{j}$ denote the position of the $jth$ vehicle; the distance between cars $%
j $ and $j+1$ is then given by $d_{j}=$ $x_{j+1}-x_{j}$. We propose
dynamical rules involving three quantities: the velocity $v$, the braking
parameter $R$ and the intercar distance $d_{j}$. In order to obtain the
state of the system at time $t+1$ (which we denote by primed quantities)
from the state at time $t$, the following four rules are applied to all cars:

\begin{enumerate}
\item  Acceleration ($\mathcal{A}$): Each vehicle increases its current
velocity if $v_{j}<v_{max}$: 
\begin{equation}
v_{j}^{\prime }\leftarrow \min (v_{j}+1,v_{\max })
\end{equation}

\item  Noise ($\mathcal{N}$): If ($v_{j}>v_{min}$), the speed of the $jth$
vehicle is reduced randomly by one with probability $R$, i.e.: 
\begin{equation}
v_{j}^{\prime }\leftarrow \max (v_{\min },v_{j}-1),
\end{equation}

\item  Proximity ($\mathcal{P}$) (to avoid collisions with other vehicles):
\ The $jth$ vehicle decelerates if it is in danger of colliding with the
vehicle $j+1$ in front of it at time $t+1$ . This represents\textit{\
anticipatory driving}:

\begin{equation}
v_{j}^{\prime }\leftarrow \min (v_{j},d_{j}+v_{j+1})
\end{equation}

\item  Movement ($\mathcal{M}$): Each vehicle is moved forward so that
\end{enumerate}

\begin{equation}
x_{j}^{\prime }=x_{j}+v_{j}
\end{equation}

The principal modification that we make is thus in step $\mathcal{D}$; in \
our model, each car has \textit{prior knowledge} of the velocities of cars
ahead of it, leading to a smoother, faster flow of traffic than that which
is simply based on a rule such as \cite{nagel}, \cite{ss} $v_{j}=\min
(v_{j},d_{j}).$

Step $\mathcal{A}$ reflects the tendency of drivers to drive as fast as
possible, without exceeding the maximum speed limit. The noise in the step $%
\mathcal{R}$ takes into account stochastic braking to do with either road
obstacles or individual reaction times of drivers; such obstacles, extrinsic
or intrinsic, often result in the spontaneous formation of a traffic jam.
The step $\mathcal{P}$ implies that the driver of a car anticipates the
position of the car in front of it at the next time step; if a collision
looks likely, the driver brakes, but not otherwise. In other words, the
driver would like to be at the maximal possible velocity consistent with the
avoidance of collisions, as in automated traffic\cite{ril}.

We apply the rules in the order $\mathcal{NPMA}$. Our initial investigations
indicated that the order $\mathcal{PNMA}$\emph{\ }led to several unphysical
configurations, whereas $\mathcal{NPMA}$\emph{\ }did not. The reason for
this is that with the noise being applied \emph{after} the proximity step,
cars were unable to adjust to the noise-reduced velocities of the traffic in
front. This frequently led to an \emph{artificial} jam, arising from the
order of the rules rather than from the real dynamics of the system. Also,
importantly, our choice of rules could be said to model the behavior of 
\textit{anticipatory} drivers rather than, as in the case of the $\mathcal{%
PNMA}$ ordering, \textit{reactive} drivers. This choice reflects our wish to
model automated cars\cite{ril}.

\section{The steady state in an open system: results and analysis}

In this section, we describe both qualitative and quantitative features of
our results for the steady state of traffic in an open system, as described
by the model presented above. We have performed extensive simulations of
this, starting with different random initial conditions, different values
for initial density $\rho _{ini}$, initial average velocity $\langle
v_{ini}\rangle $ and the braking parameter $R.$ Specifically, we have
updated individual car velocities and positions in accord with the rules
above. Regarding the boundaries, at each timestep, if the first position $%
x_{1}$ is empty, a vehicle is introduced with the maximum speed;
correspondingly, vehicles travelling beyond\ $x_{L}$, are removed. After a
transient period that depends on $L$, $R$, and on the (random) initial
conditions, the system reaches its asymptotic steady state. We present in
this paper only data obtained for system sizes $L=400,1000$ and $10,000.$

\subsection{Results}

In this subsection, we call attention to the three fundamental aspects of
our results, which we will discuss in greater detail in the following. The
first deals with the difference between the consequence of our rules and the
NaSch rules in a \textit{closed} system. The second concerns the collapse of
the fundamental diagram, which is a \textit{triangle} in the \textit{closed}
system\cite{ss}, to a \textit{point}, in an \textit{open} system. \textit{In
other words, the nature of traffic flow on average is dominated not by the
mean flux, but by the mean velocity or density - this is specific to traffic
flow in open systems}. The third is really the central result of this paper.
We find that the state of traffic in our system is determined by a)an
extrinsic parameter, which turns out to be $R$ and b)an intrinsic parameter,
related to the density $<\rho >=N/L$. This is analogous to models of
vibrated granular media\cite{amdyn}; we suggest the use of two temperatures
to describe fully the flow characteristics, and present initial evidence of
this in terms of velocity correlation functions.

\subsection{Open and closed systems: spacetime diagrams}

Fig. 1 shows the difference engendered by our specific choice of rules with
respect to the canonical NaSch model, in a \textit{closed} system. \ We plot
flux $q$ vs. density $\rho $ for a system of length $L=10000$, and for a
braking parameter $R=0.4$. Our findings are:

\begin{itemize}
\item  that the maximal value, $q_{\max }$ of flux reached is enhanced when
our specific rules are used. This is intuitively reasonable, since
anticipatory driving would encourage the efficient throughput of cars on a
highway.

\item  The highway capacity, which is defined as the density $\rho _{\max }$
at which $q_{\max }$ is attained, is \textit{also} increased in our model of
automated vehicles, with respect to the NaSch model. This too is in accord
with intuition, and is a pointer to the advantage of the use of automated
vehicles in practice.
\end{itemize}

In the following, we compare first the spacetime diagrams for open and
closed systems with our rules, and next, spacetime diagrams corresponding to
different regions of parameter space in an open system with our rules.

Fig. 2 is the spacetime diagram for a system with periodic boundary
conditions, with $R=0.7$, $\rho _{ini}=0.2$ and $L=400$, whereas Fig. 3 is
the corresponding diagram for the same parameter set in an open system. We
see that the closed system preserves its initial density, since cars cannot
be 'lost' in a system with closed boundaries, but in the case of the open
system, there is a transition from an initially low-density configuration to
a jammed steady state. Thus, the open system 'chooses' its own final
density, while the closed system simply maintains its initially chosen one.
This is reinforced by Fig. 4, which is a spacetime diagram for an open
system with $R=0.5$, $\rho _{ini}=0.7$ and $L=400$. Here, in contrast to
Fig. 3, an initially high-density system undergoes a transition to free flow
in the steady-state, when it is characterised by a much lower value of the
final density; we observe successive stages in the temporal evolution of
traffic, with the formation, coarsening and eventual dissolution of jammed
clusters.

We emphasise that in Figs. 3 and 4, the flux is a constant in the steady
state; the \ rules of our model are such that strictly one car is introduced
at the open left-hand end of the system ($x_{1}$ ), and, once the steady
state is attained, one car leaves the right-hand end ($x_{L}$) of the
system. Despite this, the nature of the flow is totally different in Figs. 3
and 4, the former indicating a transition from a free to a jammed state, and
the latter the reverse. This already demonstrates that for an open system,
the net flux is not a good descriptor of the traffic, an issue which we will
address in greater detail in the next subsection.

\subsection{The open system: flux, transients and the fundamental diagram}

It is well known that for systems with periodic boundary conditions, the
fundamental diagram of traffic flow is a triangle \cite{pr}. \ For an open
system, however, this collapses to a point, in the sense that the average
density $<\rho >$ is always inversely proportional to the average velocity $%
<v>$ \textit{in the steady state}. Our reasoning is as follows. In our
specific system, $q_{in}$, the incoming flux is strictly 1, while the
outgoing flux $q_{out}$, in the steady state, will be constrained to be 1,
thus implying an inverse relationship of velocity to density (since the net
flux is a constant). However, even if the introduction of cars at the open
left-hand end were to be more random, our argument would hold \textit{on
average}; in this case, whatever the average value $\left\langle
q_{in}\right\rangle $ of the incoming flux, the average value $\left\langle
q_{out}\right\rangle $ of the outgoing flux would be constrained to be the
same in the steady state. Thus, once again,\textit{\ in the steady state }we
would see that the average density $<\rho >$ would be inversely proportional
to the average velocity $<v>$. Fig. 5 illustrates this relationship, for a
range of values of the braking parameter $R$. Note that the full line
joining the points is meant as a guide to the eye, since in the intermediate
regions, the system exhibits 'glassy' behaviour at least for finite system
sizes (see below); this makes it very time-consuming to get systematic data
within computationally reasonable times. However, our argument above is
independent of such finite size effects, and is perfectly general.

All the above statements were made for the steady-state behaviour of our
system. However, we have also \cite{mariaelena} carried out detailed
investigations of \textit{transient} states in the open system, some
examples of which are shown in Fig. 6. These figures are plots of the flux $%
q $ versus time $t$, for different values of the braking parameter $R$; in
each plot, the different curves correspond to specific initial average
velocities $<v_{ini}>$, as indicated.

\begin{itemize}
\item  We note first of all, that each of these systems asymptotes to the
steady state where the flux is one (see above).

\item  A detailed investigation of all the simulations corresponding to the
various curves in the figures allows us to make the following non-trivial
observation: \textit{all the curves which show an initial peak (i.e. where
the value of the flux is greater than 1) are precisely those where the final
steady state corresponds to free flow, whereas all curves which show an
initial dip with a value of flux less than 1, are those where the final
steady state corresponds to jammed traffic}.

\item  A combination of the above two arguments indicates strongly that flux
is not a good descriptor of traffic in an open system, and that one has to
deal with other descriptors (for example, average densities or velocities)
to characterise the steady state of the system.

\item  We note that, as might be expected, increasing $R$ leads to far fewer
cases of free flow in the steady state.

\item  Finally we note that, for a given value of $R$, it is only the
highest initial (average) velocities that manage to make the transition to
free traffic. This shows already a strong hysteresis in the system. A simple
picture to describe this situation is the following: for each value of the
initial velocity indicated, the system is started with a fixed initial
density. The peaks (and dips) are indicators of collective events, where the
cars at a certain time cluster around selected values of velocities and
densities, \textit{during} the transient state; for example, a peak would
indicate a 'rush' of cars, which eventually could 'un-jam' the system from
an initially jammed state, in order eventually to make the transition to
free flow. Conversely, all the dips correspond to the onset of a collective
congestion, from which the system never recovers. We shall return to this
point in the last section, but we already see indications of the strong
influence of transients on steady-state traffic. This pinpoints the need to
have an appropriate descriptor mirroring the initial conditions, in addition
to the braking parameter $R$, in order to predict the steady-state behaviour
of our traffic system.
\end{itemize}

In conclusion, we summarise the two main results of this section. The first
is, that contrary to current ideas, flux cannot be used to characterise the
state of traffic in an open system; high/low values of flux are commonly
used to denote free/jammed traffic\cite{pr}, but we have shown here that for
the \textit{same} values of flux, we can have either jammed or free traffic
in an open system. The second result concerns the very interesting light
that transient behaviour can throw on the steady-state behaviour of our
traffic model (Fig. 6); this strong hysteretic behaviour is reminiscent of
granular flow, where the dynamics that take place during the formation of a
granular bed determine its subsequent structure, as well as any resulting
dynamics\cite{bm}. In the next section, we will further explore these
analogies.

\subsection{The nature of the transition as a function of $\protect\rho $
and R: analogies with granular flow}

The behaviour of traffic transients presented in Fig. 6 showed that the
initial velocity was a strong determinant of the steady state of the traffic
in the sense that collective behaviour manifested either as 'bursts' or as
'blocks' for a significant, if transient, amount of time was able to free or
jam the traffic. In Figures 7a and 7b, we explore the role of initial
densities in this regard. The average behavior of the system is determined
over an ensemble of different initial conditions: in particular, our results
are averaged on an ensemble of 100 simulations for the \emph{same} initial
density and value of $R$. Fig. 7a is a plot of density $\rho $ vs. time $t$
(on a logarithmic scale) for $L=400$, $\rho _{ini}=0.7$, for a range of
values of $R$, while Fig. 7b is the same plot for $\rho _{ini}=0.2.$ Our
finding, in the high-density case represented by Fig. 7a, is that there is a
strict phase separation between free and jammed traffic, with no
intermediate states permissible. On the contrary, in Fig, 7b, we find a
spread of final states. Even at the jammed and free ends, there is a spread
of densities; for example, in the jammed phase, the range of steady-state
densities achieved by our system ranges from $\sim 0.85$ to $\sim 0.95$.
However, and more interestingly, there is an intermediate region which does
not reach the steady state even at arbitrarily long simulation times. In
particular the data around $R=0.51$ suggests that the density grows
approximately logarithmically with time, an indication of 'glassy' behaviour
at least for this system size.

To see the extent of finite size effects in this system, we performed checks
for $L=400,1000$ and $10000$, with $\rho _{ini}=0.2$ as well as, for
comparison, the data for $L=400,$ with $\rho _{ini}=0.7.$ Our results are
shown in Fig. 8, and suggest that the transition between free and jammed
flow for $\rho _{ini}=0.2$ gets sharper for larger systems. It is rather
likely that in the limit of an infinitely large system, the 'glassy' states
causing the spread in this transition for low initial densities might
disappear and the curve might end up coinciding with that for high initial
densities (see Fig. 8). However, it should be borne in mind that real
highways are in fact of finite size, so that in our view the glassy states
causing the difference between the low and high $\rho _{ini}$ regimes are
not just of physical, but possibly also of real interest.

We replot, in Fig. 9, our earlier data in terms of $\frac{d\rho }{dR}$ vs. $%
R $ to show more clearly the nature of this transition for different initial
densities; Fig. 9a corresponds to $L=$ $400$,$10000$, with $\rho _{ini}=0.2$%
, while Fig. 9b is same plot for a system size of $L=$ $400$, \ but with $%
\rho _{ini}=0.7$. We note that, for the \textit{same} system size, the
transition from free to jammed traffic as a function of $R$ is smeared out
in the lower (initial) density case relative to the higher (initial) density
one; the reason for this is that there will always be low-density
configurations, even at the highest values of $R$, which will 'escape' the
jammed state and result in free traffic. (Fig. 10 illustrates this situation
for $L=400,R=0.8$, with $\rho _{ini}=0.2$.\ Each of the lines in the figure
corresponds to a run with these parameters, and we note that while most of
them end up in a jammed state, there are two runs which result in free
traffic). We however, also note that the width of the transition in the low
initial density case is greatly reduced for larger systems, although the
qualitative features appear to survive, e.g. the cusp in the curve. It
appears to us from these and other observations, that there will always be a
difference in the nature of the transition for the cases when arbitrarily
large systems of the \textit{same} size are started at low and high
densities (i.e. above and below about $0.5$).

What this makes clear is that contrary to some earlier speculations \cite
{nagatani}, traffic in finite systems cannot be characterised by a single
'temperature'-like variable; thus for example, in our case, it is not enough
just to regard $R$ as an effective temperature, which determines the state
that the system will reach at asymptotically long times. This underlines the
need to consider both the density $\rho $ and the braking parameter $R$ when
one tries to predict the asymptotic state of traffic in open systems.

The results of Figs. 8 and 9 suggest a clearly hysteretic behaviour of the
system, which we interpret as follows. In real traffic, it is well known
that\ the variability of road and weather conditions even on a single high
way would lead to a variability in braking rates. With this in mind, we
imagine starting in a jammed configuration (of density $\rho _{ini}=0.7$,
say as in Fig. 9b), on a road where the effective braking parameter has a
value of $0.8$. Suddenly the road conditions change dramatically, and the
effective braking parameter becomes $0.2$, say. We see from Fig. 8, that the
resultant value of the density of traffic would end up being around 0.2,
say. Now let us imagine a sudden return to bad weather/road conditions,
resulting in an effective braking parameter of = 0.8, as before. What Figure
9a as well as Fig. 10 indicate is that one can either remain in a state of
free traffic, or end up in a jammed state, depending clearly on the precise
positional correlations of the cars; in other words, there is \textit{a
finite probability that one will not return to the initial state}, i.e. in
this case, the one that was characterised by a density of $0.7$. This
indicates that if we were to plot a curve of $\rho $ vs. $R$, with these
variability conditions incorporated, we would expect to see a hysteresis
loop in density $\rho $, the size of which would clearly depend on the
length of road $L$ for which each value of $R$ would need to be implemented.
Current work is in progress to confirm this.

This behaviour is very reminiscent of that which obtains in granular media.
Imagine a box of sand is subjected to 'annealed cooling' \cite{ed}; that is,
it is submitted to different shaking intensities $\Gamma $ for variable
amounts of time $\Delta \tau $, such that after each time $\Delta \tau $,
there is a jump in shaking intensity $\Delta \Gamma $ to the next shaking
intensity. It is found experimentally \cite{ed} that a hysteresis loop is
traced out in terms of density, the size of which depends \cite{bm, paj} on
the specific ratio $\frac{\Delta \Gamma }{\Delta \tau }$, known as the 'ramp
rate' of the system. For small ramp rates, i.e. where each value of $\Gamma $
is traced out quasi-continuously, and where the system is allowed to
'equilibrate' \ at each value of the shaking intensity, the hysteresis loop
is much smaller than at larger ramp rates. The same would obtain in our
traffic flow system if we made the following analogies:

\begin{center}
\begin{tabular}{|l|l|}
\hline\hline
\textbf{Granular Flow} & \textbf{Traffic Flow} \\ \hline\hline
vibration intensity $\Gamma $ & inverse braking parameter$\frac{1}{R}$ \\ 
\hline\hline
waiting time $\Delta \tau $ & 'effective system size' $L$ \\ \hline\hline
density $\rho $ & density $\rho $ \\ \hline\hline
\end{tabular}
\end{center}

\bigskip

Based on this, we make a further analogy to do with effective temperatures
in the case of traffic flow. In the case of granular media, it is now
conventional to refer to the vibration intensity and density respectively as
being related to effective temperatures corresponding to the fast and slow
dynamics of this athermal and complex system; a related fast dynamics
'temperature' has been in use for a long time by the engineering community 
\cite{savage}, while a version of the slow dynamics temperature (termed the
compactivity)\ was first proposed by Edwards and collaborators\cite{sam}.
Subsequent work has confirmed the need to use both \cite{ars, amdyn} these
temperatures in any analysis of the dynamics of granular media. We here thus
propose, based on our present results:

\begin{itemize}
\item  the use of the inverse braking parameter $\frac{1}{R}$ as an
effective temperature which controls the 'fast' or 'single-car' dynamics of
traffic flow

\item  the use of the inverse density $\frac{1}{\rho }$ as an effective
temperature which controls the 'slow' or 'collective' dynamics of traffic
flow
\end{itemize}

Clearly, this analogy is relatively qualitative at this point, and in this
paper, we quantify it to a slightly greater extent with the use of velocity
correlation function. However, we emphasise that more analysis, especially
to do with the tracing out of the hysteresis curve referred to earlier, is
in progress.

In this tentative spirit, we examine equal-time velocity correlation
functions as a function of space, in the steady state, for systems of
traffic which are started out with rather different initial conditions. In.
Fig. 11a, we present averages over 200 runs on a system of size $L=400$ for
the quantity $<v(0,t)v(x,t)>$ vs space $x$ with an initial density of $\rho
_{ini}=0.7$ and different values of $R$, as designated on the legend of the
figure. In Fig. 11b, corresponding data are presented for systems with an
initial density of $\rho _{ini}=0.2$ \ and $R=0.3,$ and $0.8$.

\begin{itemize}
\item  In Fig. 11a, we notice that for $R>0.5$, there is little free volume
for the cars to move, so that the initial jammed state persists more or
less. At $R=0.5$, there is the appearance of a shell structure, which
becomes increasingly evident for progressively lower values of $R$. We note
that in fact, for $R\in \left( 0.4,0.5\right) $, the minima and the maxima
of the shells are the sharpest, with the most regularity in shell spacing;
for lower values, the structure becomes more diffuse, until at $R=0.2$, we
have a 'liquid-like' structure. This might be an indicator that the jammed
state gives way to an 'ordered' state for $R\sim 0.4$, and the ordering
disappears rapidly as we reach lower values of the braking parameter. This
is analogous to the situation in granular media:\ starting from totally
jammed configurations (typically characterised by a value of the density
corresponding to random close packing \cite{bernal}, although this has
recently been the subject of considerable debate\cite{torquato}), when the
grains have just enough free volume to move, it is known that the system
moves preferentially to configurations which have some semblance of order 
\cite{bm, amdyn}. As the excitation intensity $\Gamma $ increases, this
order gives way to a liquid-like structure, characterised by more diffuse
spatiotemporal correlation functions.

\item  In Fig. 11b, we notice a rather distinctive difference. As remarked
before, the low-density system, even at high values of the braking parameter 
$R$, has a small but finite probability to remain in the free traffic phase.
Thus a very small fraction of the runs carried out at $R=$ $0.8$ result in
the free state (cf. Fig. 10), while typically the system ends up in the
congested state; it is therefore rather difficult to say anything conclusive
about the free state at $R=$ $0.8$, but we can see rather clearly that the
jammed state in this case shows some structure, which is somewhat different
to that manifested by the jammed state in Fig. 11a. This is another
manifestation of strong hysteretic effects, referred to earlier. We
speculate a particular reason for the appearance of some shell-like
structure in this case (as contrasted with the totally jammed corresponding
configuration in Fig. 11a) may be the formation of the 'free zone'
manifested in Fig. 3 at the open end, which appears to be rather typical\cite
{mariaelena} of jammed states reached from initially low-density
configurations. When $R=$ $0.3$, the structure of the system is entirely
'liquid-like', and not distinguishable from the liquid-like state reached in
Fig. 11a (when the system was started at a high initial density). This
liquid-like structure in both cases is reminiscent of the fluidised state in
a granular medium shaken at high intensities of vibration \cite{bm, paj}.
\end{itemize}

Additionally, we remark on the specific meaning of such \emph{dynamical}
correlations; in analogy with earlier work on granular flow \cite{bm}, we
define a \emph{dynamical cluster} for a given $R$ as being the number of
sites which are within the first shell of the velocity correlation function. 
\textit{The physical import of a dynamical cluster is that it reflects the
range over which cars are correlated in their velocities. } In general, as
fewer cars face random obstacles, more and more of them develop velocity
correlations, i.e. they begin to 'move together' in clumps. Returning to the
analogy with granular flow, this mirrors the situation found in earlier work 
\cite{bm} where a \textit{decrease in external perturbations applied to a
granular system} (or in our case \textit{an increase in the braking
parameter }$R$) causes an \textit{increase} in the size of a typical
dynamical cluster of grains. We remark here that dynamical correlations of a
similar sort have also subsequently been observed in glasses, where they are
commonly referred to as spatial heterogeneities\cite{glotzer}. This
similarity of behaviour of our traffic flow model with granular and glassy
materials is an additional indicator that two temperatures are necessary to
characterise our system, since recent work in both glassy\cite{kurchan} and
granular\cite{amdyn, paj} materials has indicated the ubiquity of this
description in both systems. Further work is in progress to quantify these
analogies, especially to do with unequal time correlation functions.

\section{Discussion}

In this paper we have analysed a model representing traffic flow in a system
which involves automated cars, and hence anticipatory driving, which was
motivated by current interest on automated highways in real traffic systems 
\cite{ril}. While we have here focused on the statistical mechanical aspects
of this model, we point out that a very detailed discussion of the practical
aspects of our model, with special relevance to real automated vehicles, can
be found in\cite{mariaelena, tm}.

One of our major results concerns the difference between open and closed
traffic systems. In particular, in the former, we find that the nature of
the phase diagram is completely altered with respect to the latter; for
example, the fundamental diagram of flux versus density as a function of the
parameter $R$ presented recently for closed systems \cite{eis, pr} collapses
to a point in the case of an open system. We find thus that flux is not a
good descriptor of open systems, and that a knowledge of densities and
velocities is necessary to characterise the difference between free and
jammed traffic. This is indicated both by the steady-state, as well as the
transient behaviour of our traffic model.

Our other major result concerns the 'glassy' or first-order nature of the
dynamics of our traffic flow model in an open system of finite size. We have
shown, both by a detailed examination of the effect of varying the initial
density of the traffic system, and then subjecting it to varying braking
parameters $R$, as well as by an examination of equal-time correlation
functions, that two effective temperatures are needed to characterise the
steady state of our traffic model. As in the case of granular flow, another
example of an athermal complex system which exhibits glassy dynamics, we
suggest that the 'fast' (single-particle) dynamics temperature is related to
the external perturbation (vibration intensity in the case of granular
media, and the inverse braking parameter $\frac{1}{R}$ in our case), while
the 'slow' (collective) dynamics temperature is related in both cases to the
inverse of the density $\rho $.

Further work is in progress to examine this analogy in greater detail, in
particular to do with aspects related to the hysteresis curve referred to
above, as well as the analysis of two-time correlation functions in our
present model.

\section{Acknowledgments}

AM is very grateful for the generous hospitality, over many visits, to the
Centro de Investigaci\'{o}n en Energ\'{\i}a in Temixco, where a large
portion of this work was carried out. This work was partially supported by
DGAPA-UNAM under project IN103100. We are very grateful to Silvio Franz for
illuminating discussions, and to Mariano L\'opez de Haro for a careful
reading of our manuscript.

\newpage

\section{Figure Captions}

Figure 1. Fundamental diagrams for the closed system comparing our model
with the NaSch model, for $R=0.4$ and $L=10,000$. Note that the maximal flux
as well as the highway capacity are enhanced in our model with respect to
NaSch.

Figure 2. Spacetime diagram for traffic flow in a closed system ($L=400$)
corresponding to a braking probability $R=0.7$, and starting with an initial
density $\rho _{ini}=0.2$ ($v=1$ in red, $v=2$ in orange, $v=3$ in yellow, $%
v=4$ in green and $v=5$ in blue)$.$ The system preserves its initial density.

Figure 3. Spacetime diagram for traffic flow in an open system ($L=400$)
corresponding to a braking probability $R=0.7$, and a density $\rho =0.2$ ($%
v=1$ in red, $v=2$ in orange, $v=3$ in yellow, $v=4$ in green and $v=5$ in
blue). An initial state of free traffic gives way to a congested state.
Note, however, the persistence of a small zone of free traffic near the
left-hand boundary (see section 3).

Figure 4. Spacetime diagram for traffic flow in an open system ($L=400$)
corresponding to a braking probability $R=0.2$, and a density $\rho =0.7$.
An initially jammed state gives way to freely flowing traffic. Note the
formation, coarsening, and eventual dissolution of areas of congestion
before the steady state is reached.

Figure 5. Plots of the average density $<\rho >$ vs. the average velocity $%
<v>$ in he steady state, with different initial conditions, and $L=400$, for
different values of the braking parameter $R$ values indicated by different
symbols on the figure. Note that the solid line is meant as a guide to the
eye.

Figure 6. Plots of the time evolution of the density of traffic in an open
system for an initial density $<\rho _{ini}>=0.4$ and $L=400$. The different
figures correspond to different values of the braking parameter $R$ as
indicated by different symbols on the figure. On each figure, the
differently coloured lines correspond to different values of $<v_{ini}>$,
also indicated on the respective figures. Note that all curves which have an
initial peak result in a steady state of free traffic, whereas the reverse
is the case for all curves which manifest an initial dip, thus emphasising
the importance of transient behaviour for the steady state of traffic.

Figure 7. Plots of density $\rho $ vs. time $t$ (on a logarithmic scale) for 
$L=400$, and a) $\rho _{ini}=0.7$, and b) $\rho _{ini}=0.2$ for a range of
values of $R$. In Fig. 7a, note the strict phase separation between free and
jammed traffic, with no intermediate states. In Fig, 7b, there is a spread
of final states. In particular the data around $R=0.51$ suggests that the
density grows approximately logarithmically with time, an indication of
'glassy' behaviour at least for this system size.

Figure 8. Plots of density $\rho $ vs. braking parameter $R$ for system
sizes $L=400,1000$ and $10000$, with $\rho _{ini}=0.2$. The data for $\rho
_{ini}=0.7$ with $L=400,$ are shown for comparison.

Figure 9. Plots of $\frac{d\rho }{dR}$ vs $R$ for a)$\rho _{ini}=0.2$ and $%
L=400,10000$ b) for $\rho _{ini}=0.7$ and $L=400$. Note the qualitative
similarity between the curves in a), despite their difference in size.

Figure 10. Plots of density $\rho $ vs. time $t$ for $L=400$, and $\rho
_{ini}=0.2$. The lines correspond to different runs with the \textit{same}
initial densities but \textit{different positional correlations} between the
cars, which result in a spread of different steady states.

Figure 11. Plots of velocity-velocity correlation functions $%
<v_{x}v_{x^{\prime }}>$ \ for a)$\rho _{ini}=0.7$ b) $\rho _{ini}=0.2$ for $%
L=400$, corresponding to a range of different values of the braking
probability $R$, whose values are indicated by the legends on the plots. In
a), note the collapse of the curves for all values of $R\geq 0.6$. Note that
the 'shells' for lower values of $R$ get increasingly diffuse, until the
liquid-like structure corresponding to $R=0.2$ is reached. In b) note the
difference of the jammed state (red line) with the corresponding jammed
state (yellow line) in a), which confirms the hysteretic behaviour. The free
state reached for $R=0.3$, is, however virtually identical to the free state
reached in a), indicating a 'liquid-like' structure.


\begin{thebibliography}{99}
\bibitem{tft}  N. Gartner, H. Mahmassani, C. H. Messer, R. Cunard and A.
Rathy, Traffic Flow Theory: A State-of-Art-Report, monograph, published by
Transportation Research Board Committee on Traffic Flow Theory and
characteristic (1987).

\bibitem{pr}  D. Chowdhury, L. Santen and A. Schadschneider, Physics Reports 
\textbf{329}, 199 (2000).

\bibitem{ril}  J.H. Rillings, Automated Highways, Scientific American, 
\textbf{365}, 60-63 (1997).

\bibitem{green}  B.D. Greenshields, A Study of Traffic Capacity. Highways
Research Record 14, 448 (1934).

\bibitem{adams}  W.F. Adams, Road traffic considered as a Random Series, J.
Inst. Civil. Engin., London, 1936.

\bibitem{lighthill}  M.J. Lighthill and G.B. Withman, On kinematics waves
part. II, Procc. Royal Soc., Series A Mathematical and Physical Sciences, N.
1178, V. 229, London, 1955.

\bibitem{prigogine}  I. Prigogine and F.C. Andrews, Oper. Res. 8, 789
(1960); I. Prigogine and R. Herman, \textit{Kinetic Theory of Vehicular
Traffic}, Elsevier, N.Y. (1971).

\bibitem{jsp}  H. Reiss, A.D. Hammerich and E.W. Montroll, J. Stat. Phys.
42, 647 (1986).

\bibitem{chandler}  R. E. Chandler, R. Herman, and E.W. Montroll Oper. Res.
6, 165 (1958). D.C. Gazis, R. Herman, R.B. Potts, Oper. Res. 7, 499 (1959).

\bibitem{nagelpre}  K. Nagel, Phys. Rev. E 53, 4655 (1996).

\bibitem{nagel}  K. Nagel and M. Schreckenberg, J. Phys. I (France) \textbf{%
I2}, 2221 (1992).

\bibitem{leutzbach}  W. Leutzbach, Some remarks on the hystory of the
science of traffic flow in \textit{Traffic and Granular Flow}, Eds. D.E.
Wolf, M. Schreckenberg and A. Bachem World Scientific, Singapore, 1996.

\bibitem{wolf}  D.E. Wolf, M. Schreckenberg and A. Bachem, \textit{Traffic
and Granular Flow, World Scientific}, Singapore, 1996.

\bibitem{book98}  M. Schreckenberg and D.E. Wolf, \textit{Traffic and
Granular Flow'97}, Springer Singapore, 1998.

\bibitem{book}  D. Helbing, H. J. Hermann, M. Schreckenberg and D.\ E. Wolf
Traffic and Granular Flow, Springer-Verlag, Berlin (2000).

\bibitem{paj}  P. F. Stadler, Anita Mehta and J. M. Luck, condmat 0103076.

\bibitem{mariaelena}  M. E. L\'{a}rraga, ''\textit{Un Aut\'{o}mata Celular
Probabilista para la Simualci\'{o}n del Tr\'{a}nsito de Autom\'{o}viles
Automatizados}'', Master's Degree Thesis, Universidad Nacional Aut\'{o}noma
de M\'{e}xico (M\'{e}xico, 2001).

\bibitem{ss}  A. Schadschneider and M. Schreckenberg, J. Phys. A \textbf{26}%
, L679 (1993).

\bibitem{eis}  L. Eisenbl\"{a}tter, L. Santen, A Schadschneider and M.
Schreckenberg, Phys. Rev. E \textbf{57}, 1309 (1998).

\bibitem{snagelreview}  H. M. Jaeger, S. R. Nagel and R. P. Behringer, Rev.
Mod. Phys. \textbf{68}, 1259 (1996).

\bibitem{sam}  S. F. Edwards, in \textit{Granular Matter: An
Interdisciplinary Approach}, ed. Anita Mehta (Springer-Verlag, New York,
1994).

\bibitem{mybook}  Anita Mehta, Granular Matter: An Interdisciplinary
Approach. Springer-Verlag (1994).

\bibitem{bm}  G. C. Barker and Anita Mehta, Phys. Rev. A \textbf{45}, 3435
(1992); Anita Mehta and G. C. Barker Phys. Rev. Lett. \textbf{67}, 394
(1991); G C\ Barker and Anita Mehta, cond-mat/0010268.

\bibitem{santen}  C. Appert, L. Santen,Phys. Rev. Lett. \textbf{86}, 2498
(2001).

\bibitem{amdyn}  Anita Mehta and G. C. Barker, J. Phys.: Condens.Matter 
\textbf{12} (2000) 6619-6628; J. M.Berg and Anita Mehta cond-mat/0012416
(2000).

\bibitem{nagatani}  T. Nagatani, J. Phys. A \textbf{28}, L119 (1995).

\bibitem{ed}  E. R. Nowak, J. Knight, E. Ben-Naim, H. M. Jaeger and S. R.
Nagel, Phys. Rev. E \textbf{57 }, 1971 (1998).

\bibitem{savage}  S. B. Savage Adv. Appl. Mech., \textbf{24} 289 (1984).

\bibitem{ars}  Anita Mehta, R. J. Needs and Sushanta Dattagupta, J. Stat.
Phys. \textbf{68} 1131 (1992).

\bibitem{bernal}  J.D., Bernal Nature \textbf{183} 141 (1959).

\bibitem{torquato}  S, Torquato, TM, Truskett, PG Debenedetti, Phys. Rev.
Let.t 84, 2064 (2000).

\bibitem{glotzer}  P.H., Poole, C., Donati, S.C. Glotzer, Physica \textbf{A
261}, 51, (1998).

\bibitem{kurchan}  Kurchan J, J. Phys-Condens. Mat \textbf{12}, 6611,
(2000); L. Berthier, L. Cugliandolo, and J. L. Iguain, Phys. Rev. \textbf{E
63}, 051302 (2001).

\bibitem{tm}  M. E. L\'{a}rraga and J. A. del R\'{i}o, submitted to
Transport. Res. \textbf{B }(2001).
\end{thebibliography}
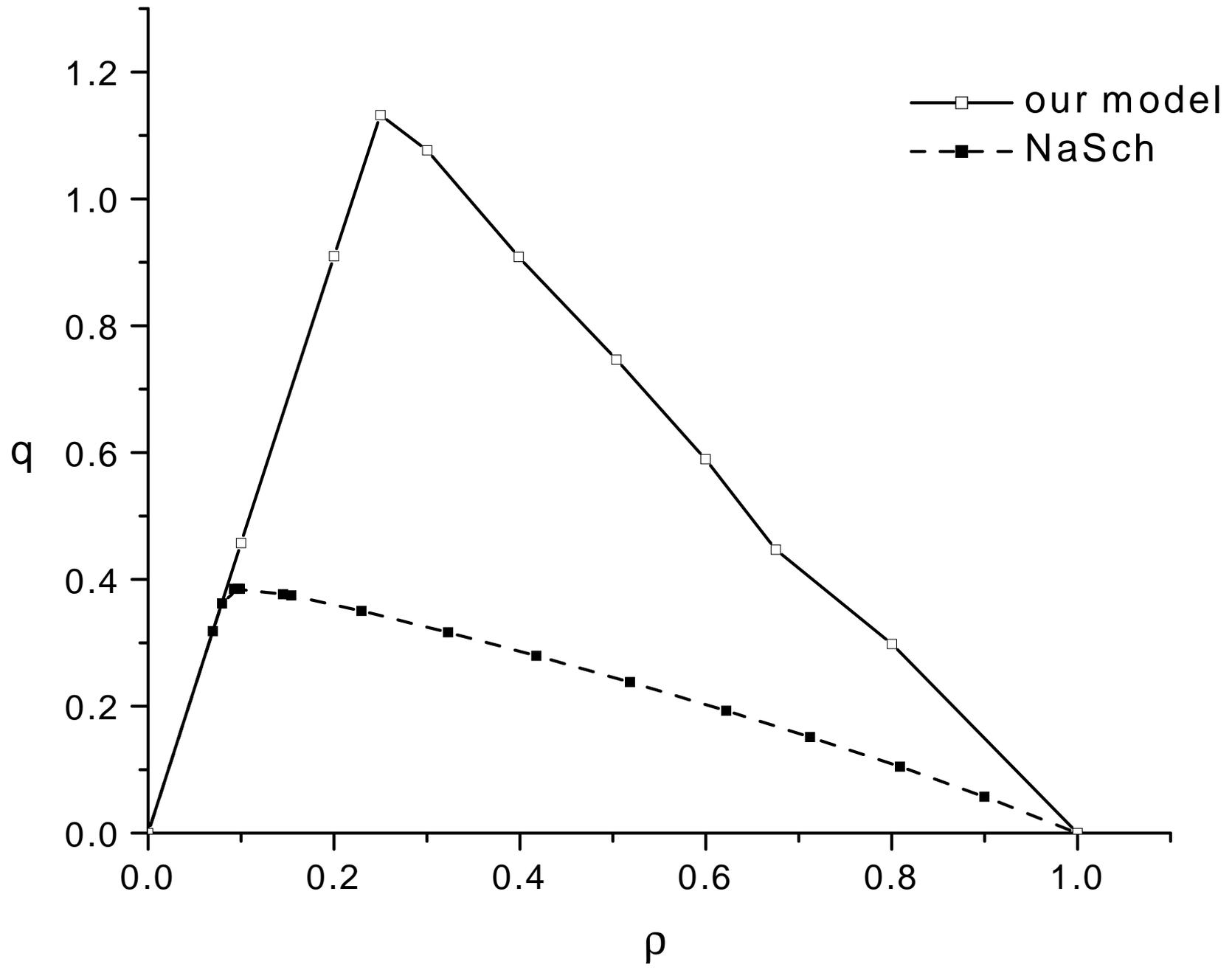
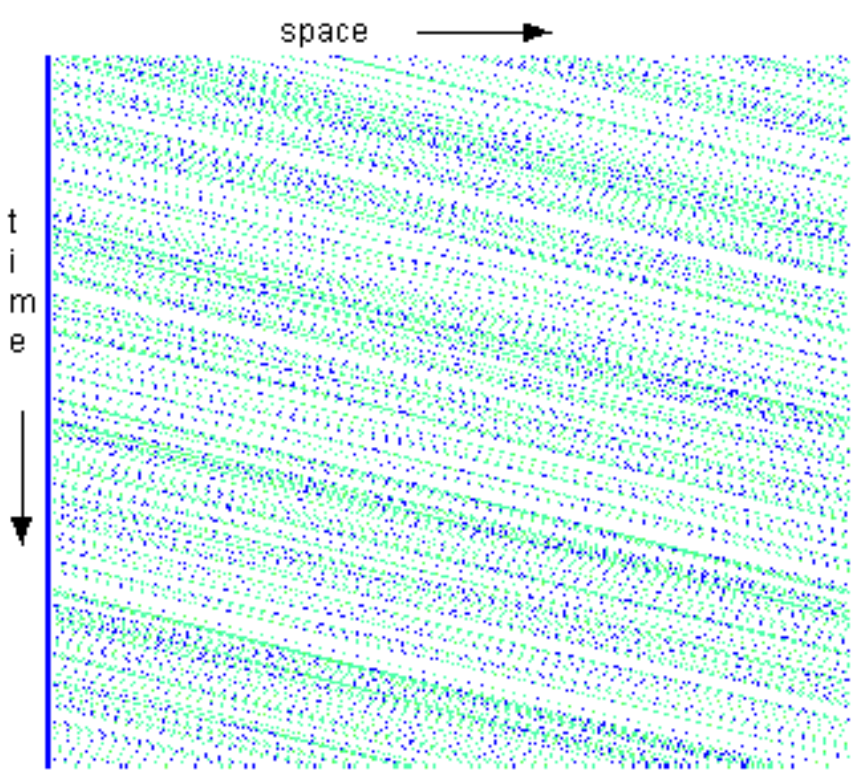
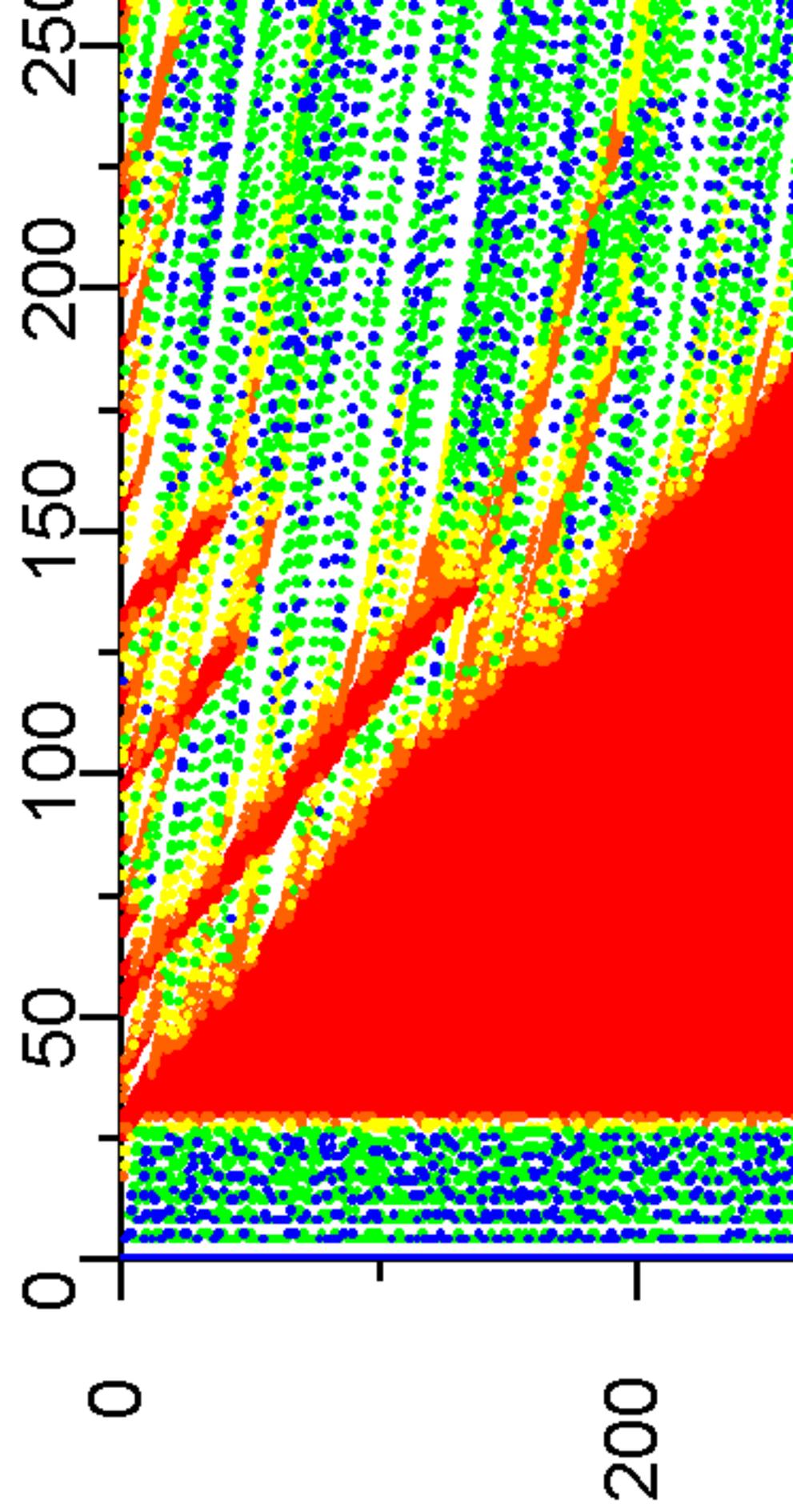
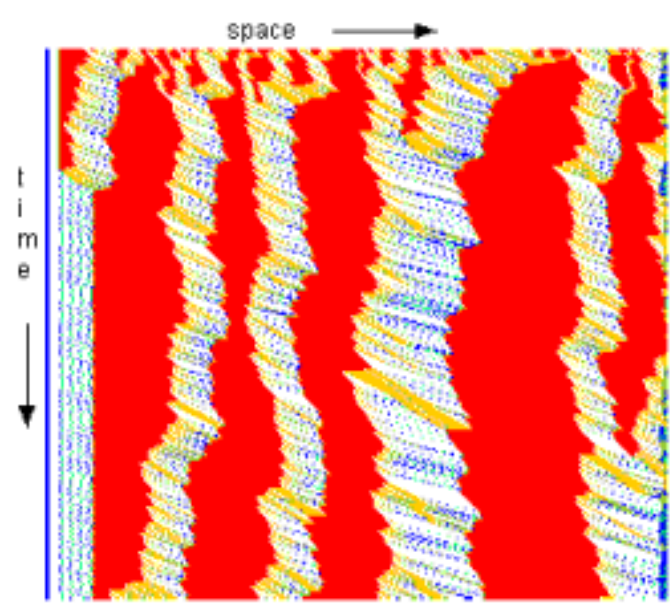
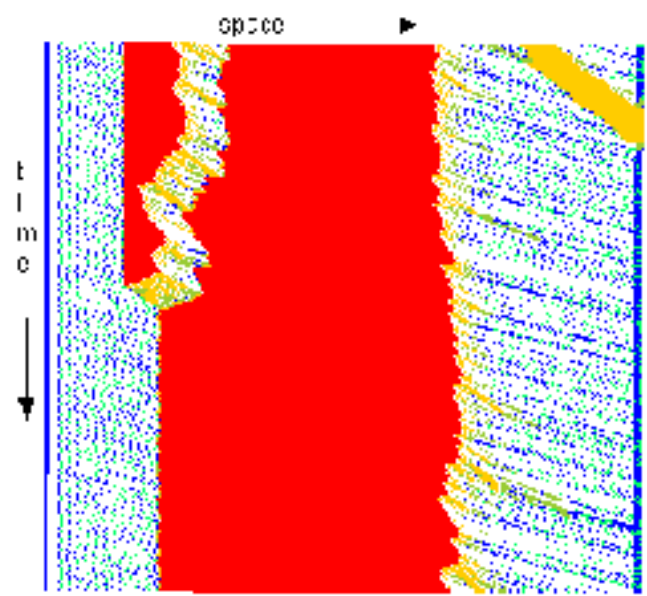
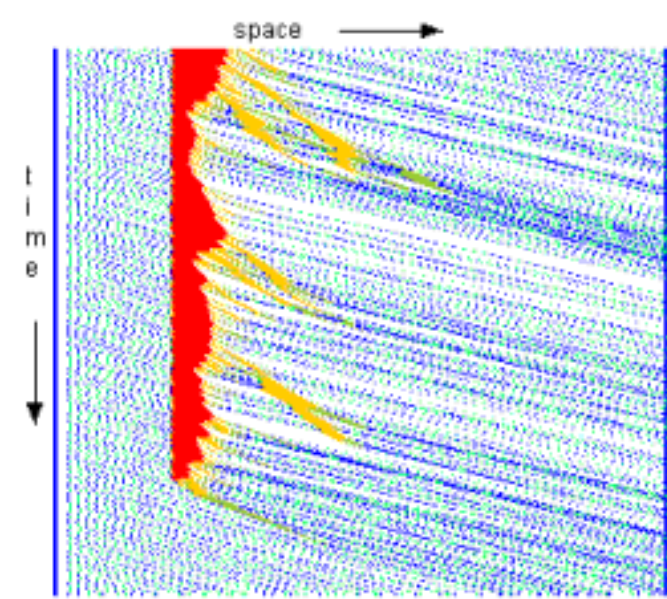
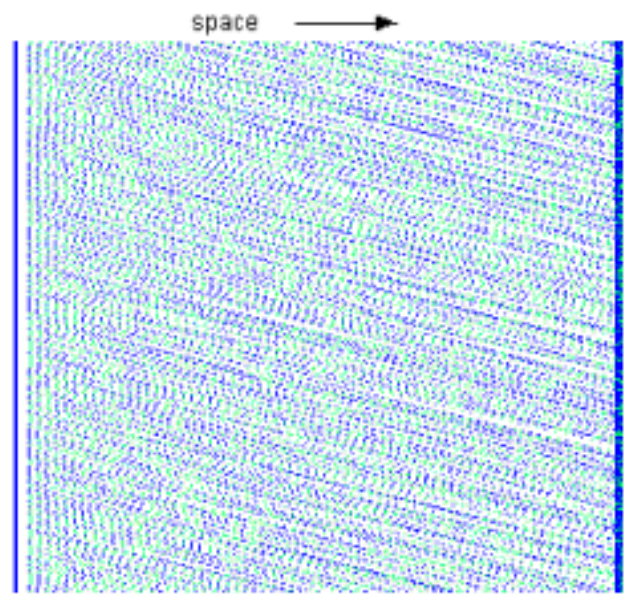
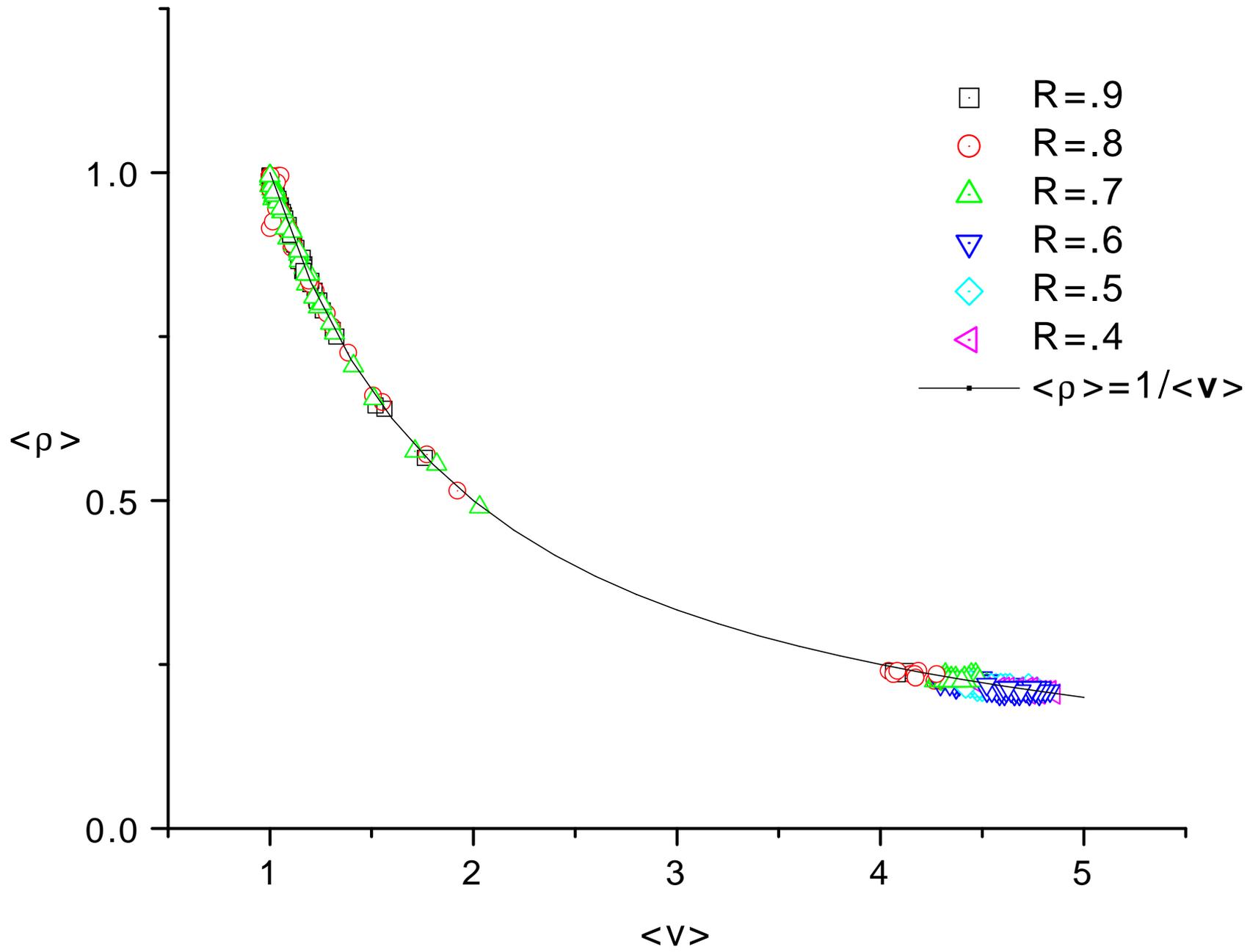
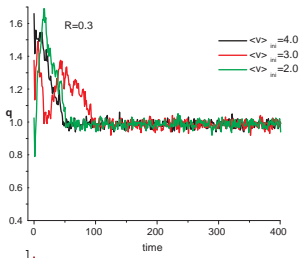
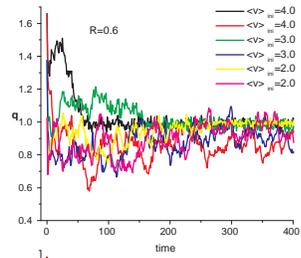
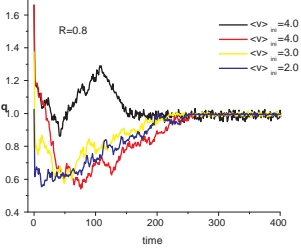
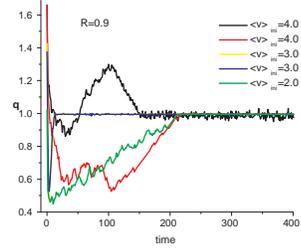
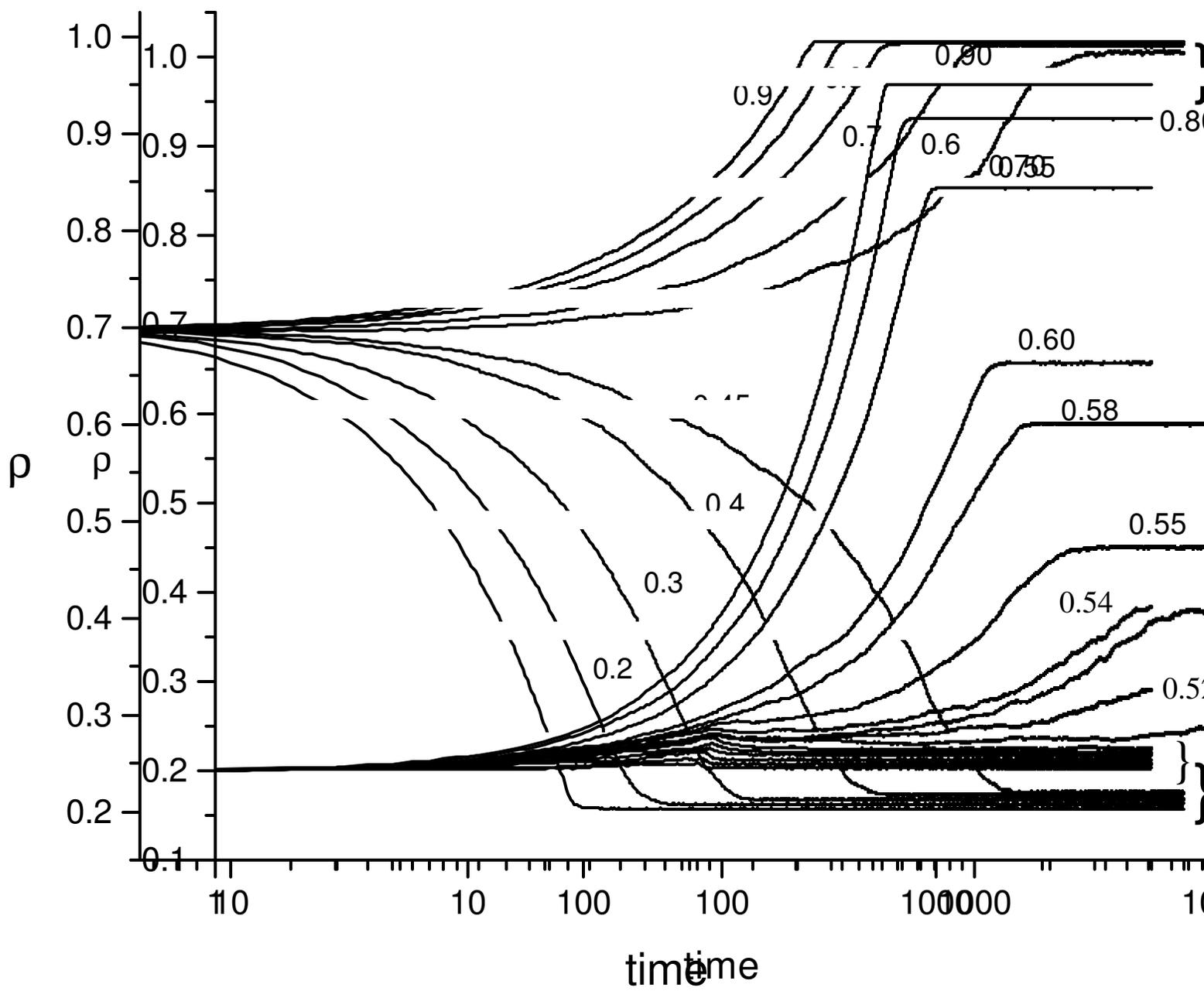
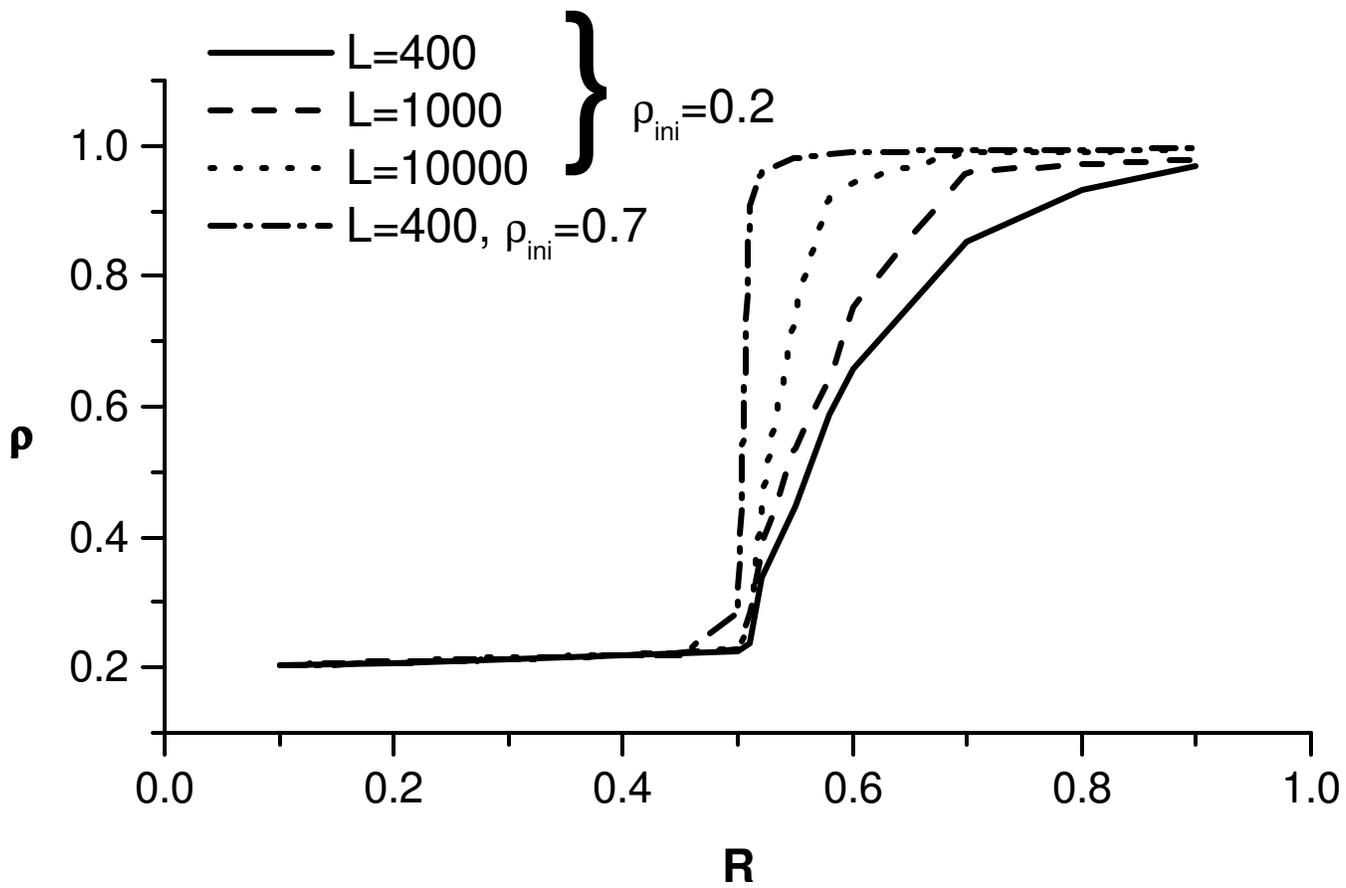
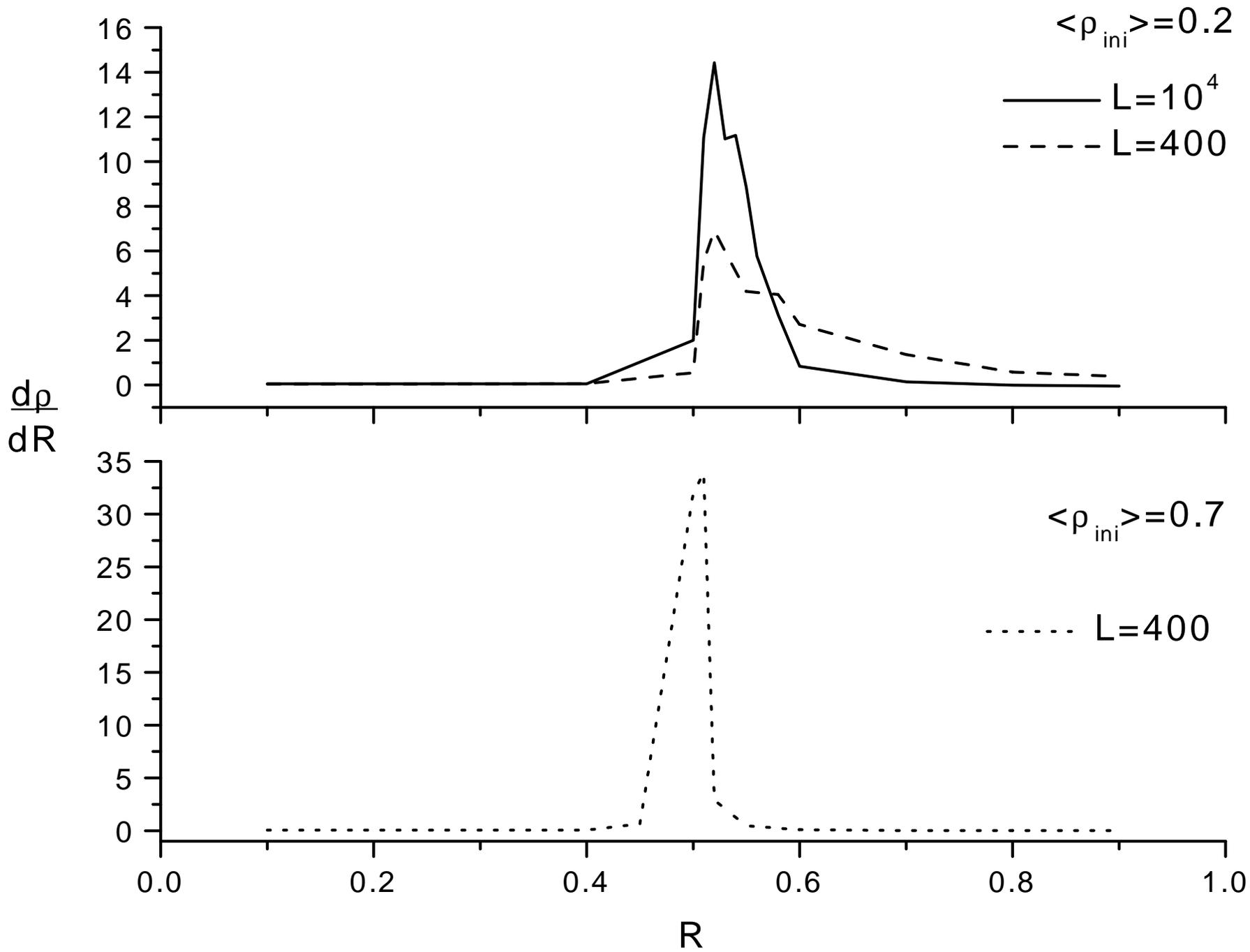
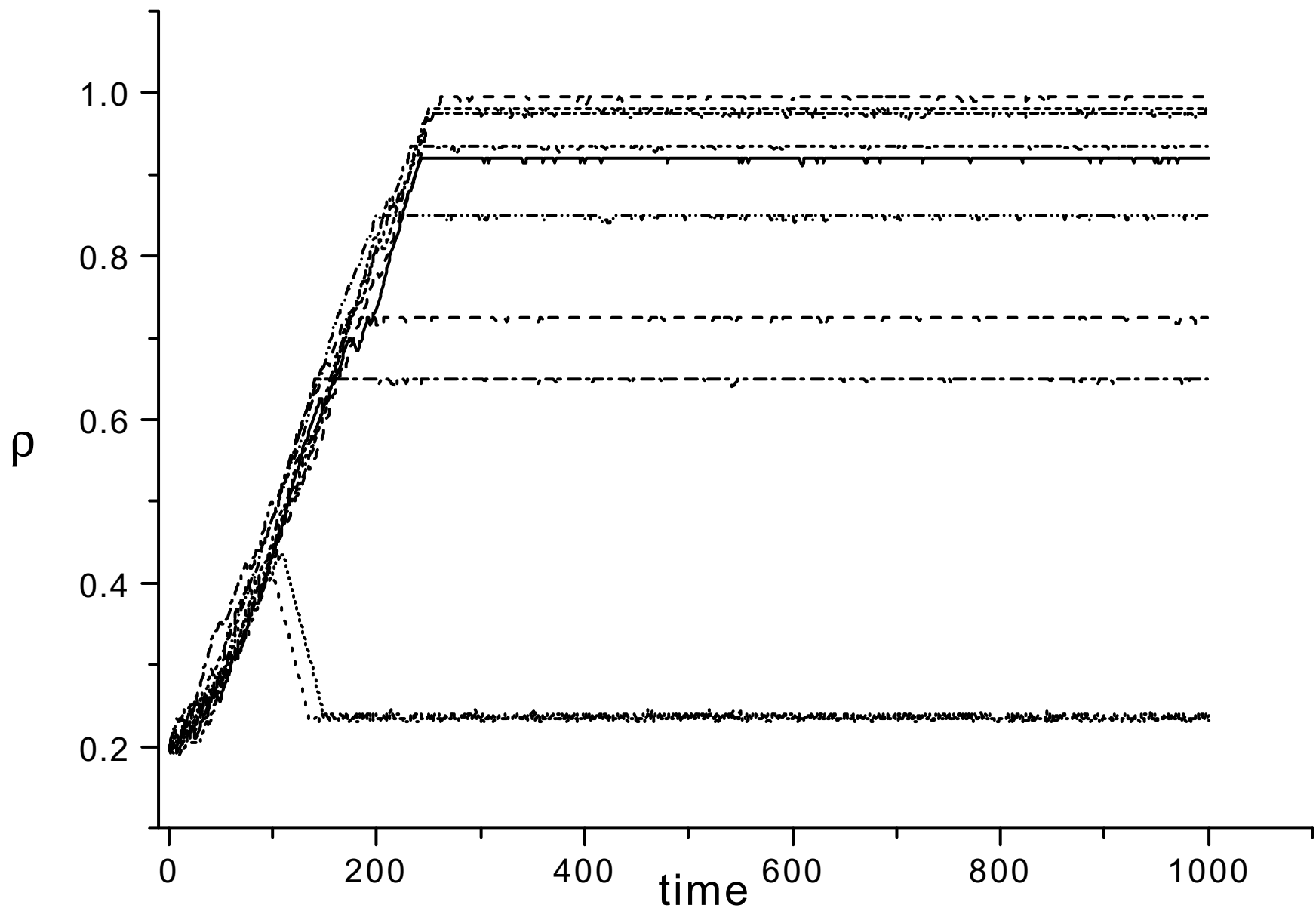
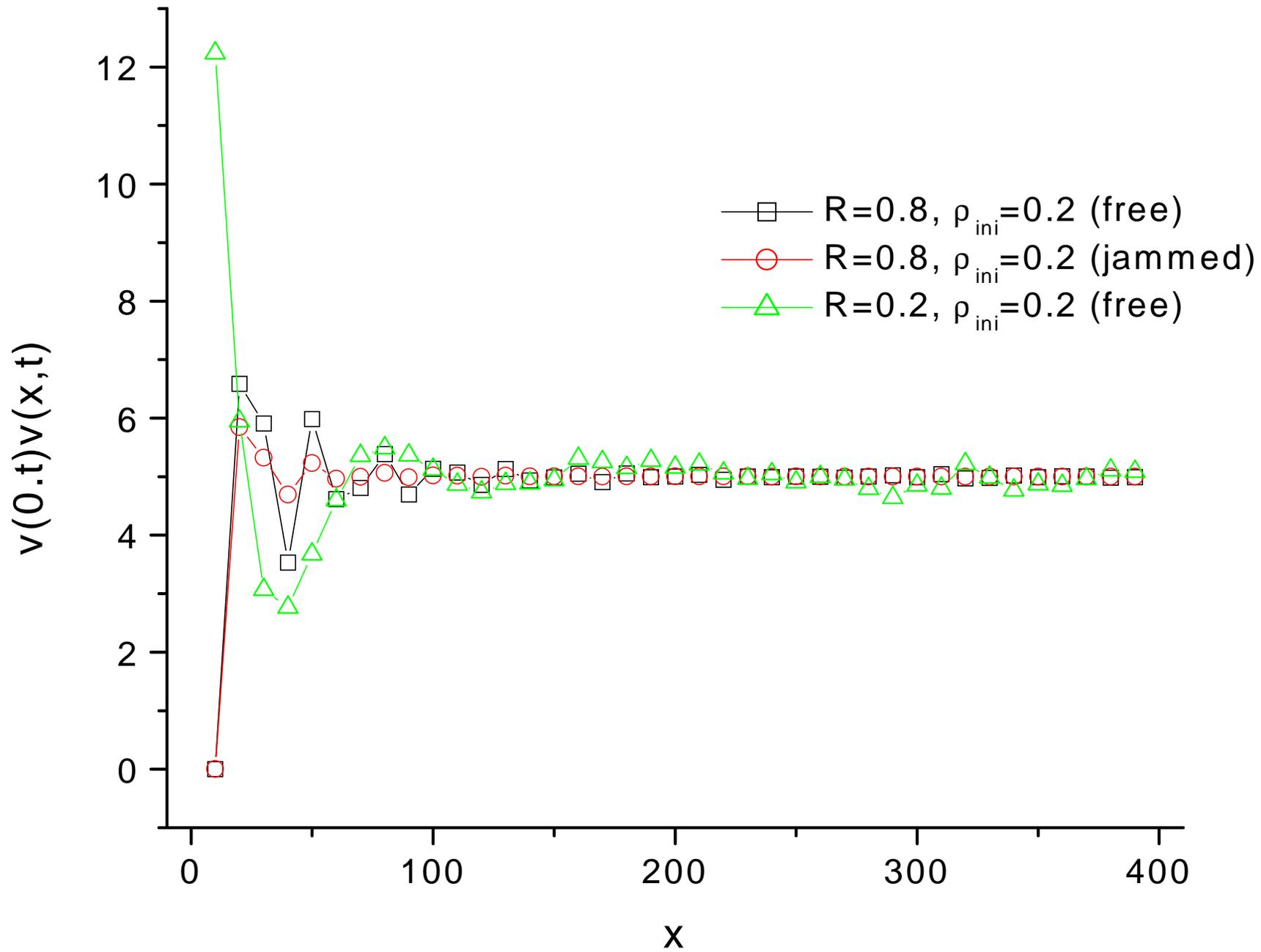
\end{document}